# DSCOT: AN NFT-BASED BLOCKCHAIN ARCHITECTURE FOR THE AUTHENTICATION OF IOT-BASED SMART DEVICES IN SMART CITIES

Usman Khalil[1] , Owais Ahmed Malik[1,2] , Ong Wee Hong[1] , Mueen Uddin[3] (Sr. Member IEEE)

**Abstract**

Smart city architecture brings all the underlying architectures, i.e., Internet of Things (IoT), Cyber-Physical Systems (CPSs), Internet of Cyber-Physical Things (IoCPT), and Internet of Everything (IoE), together to work as a system under its umbrella. The goal of smart city architecture is to come up with a solution that may integrate all the real-time response applications. However, the cyber-physical space poses threats that can jeopardize the working of a smart city where all the data belonging to people, systems, and processes will be at risk. Various architectures based on centralized and distributed mechanisms support smart cities; however, the security concerns regarding traceability, scalability, security services, platform assistance, and resource management persist. In this paper, private blockchain-based architecture Decentralized Smart City of Things (DSCoT) is proposed. It actively utilizes fog computing for all the users and smart devices connected to a fog node in a particular management system in a smart city, i.e., a smart house or hospital, etc. Non-fungible tokens (NFTs) have been utilized for representation to define smart device attributes. NFTs are unique and non-interchangeable units of data stored on a digital ledger and are widely used in blockchain-based solutions to represent unique assets. NFTs in the proposed DSCoT architecture provide devices and user authentication (IoT) functionality. DSCoT has been designed to provide a smart city solution that ensures robust security features such as Confidentiality, Integrity, Availability (CIA), and authorization by defining new attributes and functions for Owner, User, Fog, and IoT devices authentication. The evaluation of the proposed functions and components in terms of Gas consumption and time complexity has shown promising results. Comparatively, the Gas consumption for minting DSCoT NFT showed approximately 27%, and a DSCoT *approve()* was approximately 11% more efficient than the PUF-based NFT solution.

## 1 INTRODUCTION

The working mechanics of smart components in terms of people, processes, data, and things play specific roles and work together to enable future cities and communities to give rise to the concept of smart cities [1]. Smart city architecture brings together all the underlying architectures, i.e., Internet of Things (IoT), Cyber-Physical Systems (CPSs), Internet of Cyber-Physical Things (IoCPT), and Internet of Everything (IoE), together to work as a system under its umbrella. The devices associated with these architectures connect to the internet to generate, receive and process the data to make industries, healthcare, and cars more intelligent and efficient. More complex large-scale systems have been developed and deployed at the industry level to safeguard the privacy and security of cyber-physical systems (CPSs) [2].

| | Usman Khalil |
|---|---|
| ✉ | uskhalil@gmail.com, 19h8340@ubd.edu.bn |
| 1 | School of Digital Science, Universiti Brunei Darussalam, Jalan Tungku Link, Gadong BE1410, Brunei Darussalam |
| 2 | Institute of Applied Data Analytics, Universiti Brunei Darussalam, Jalan Tungku Link, Gadong BE1410, Brunei Darussalam |
| 3 | College of Computing and Information Technology University for Science and Technology, Doha, 24449, Qatar |

One such popular example is the Supervisory Control and Data Acquisition system (SCADA) [1], [3], [4]. The IoT networks deployed in different physical systems further distribute the data to the cloud, fog, and edge layer for processing at different levels following the Internet of Things (IoT) paradigms.

Figure 1 presents the smart city's generalized architecture, depicting different CPSs working in different domains such as smart homes, smart grids, smart health monitoring, smart vehicles (UAVs – Unmanned Air Vehicles, UGVs - Unmanned Ground Vehicles), process control, oil, and gas distribution, transportation systems, etc. It utilizes cloud computing as a platform-based service model for data access, storage, analysis, and network to centralized data centers and IP networks. In a smart cities concept, these CPSs are managed by national and private organizations that work along with government bodies such as municipal committees. The municipal departments manage the operations through the municipal command and control center, also known as Security Operations Center (SOC). This center connects to the internet to deploy the functionality using cloud platforms and services (i.e., cloud services, cloud storage services, and cloud management services) [5], as shown in Figure 1.

CPSs heavily depend on the edge of the network that contains the edge nodes. These edge nodes provide critical information in a physical environment within the cyberspace. Since these devices are low-powered with limited resources regarding their data collection, storage

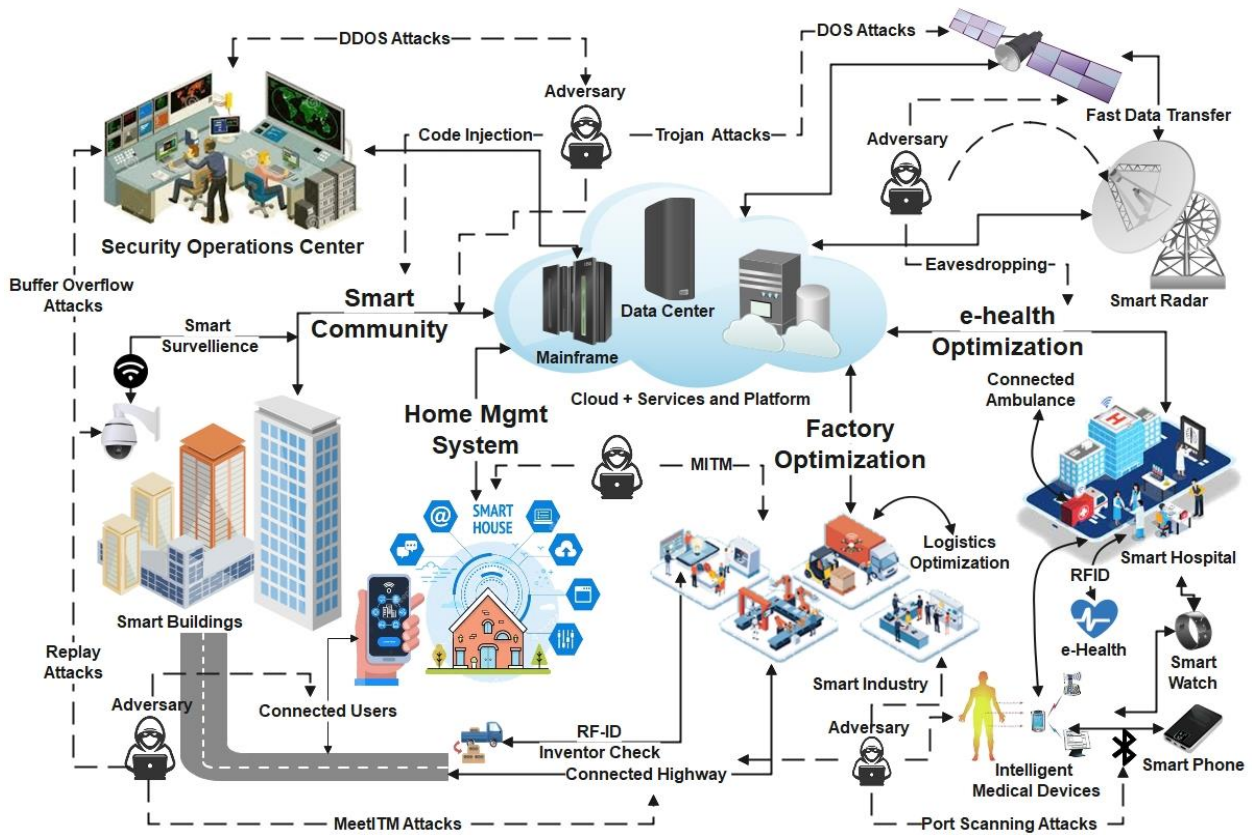

Figure 1. Generalized smart city architecture and the attack vector

, and processing efficiency, the IoT networks in CPSs have been an open playing ground for attackers, as shown in Figure 1. Also, smart cities utilize technologies like software-defined networking (SDN), cloud computing (CC), and fog computing, inheriting the current threats in those arenas [6].

### 1.1 Security and Authentication Issues of IoT-enabled Smart Devices in Smart City

Specifically, in the context of smart cities and an increasing number of IoT-enabled smart devices connecting to the internet daily, the security and authentication of these devices have become inevitable. One of the recently carried out Cisco Annual Internet surveys (2018–2023) projected internet connectivity to nearly two-thirds of the global population by 2023, showing a significant number of devices that might connect to the internet in the future [7].

Since the internet uses the transmission control protocol/Internet protocol (TCP/IP), the underlying architectures of the smart city inherit the IoT-enabled smart devices' security and authentication issues and the networks themselves [6]. Other factors include the manufacturers' low concentration in security features of the customer premises equipment (CPE), such as weak versions of SSL (v2, v3, and CBC mode) services, easily guessable default login credentials, unguarded ports, unencrypted and self-signed or expired security certificates, etc. The manufacturers of these devices left unattended authentication and access control schemes which increases the chance of exploitation in the internet infrastructures, industrial CPSs, healthcare agriculture, supply chain business, etc., as depicted in Figure 1.

Thus, a need to develop secure architectures is inevitable that may cope with the security and authentication issues of IoT-enabled smart devices operating in the underlying smart city architecture.

### 1.2 Blockchain Tokenization

Blockchain-based tokenization presents an opportunity for asset identification and authentication schemes in smart city architecture. After a huge appreciation of Token creation in 2018, with over 1,132 ICOs and STOs collected, nearly $20bn [8], the concept of tokens has gained wide attention. Tokenization in BC presents the concept of digital representation of an asset on the blockchain or colloquially "programmable asset". BC tokenization presents different types of tokens, tangible or intangible, as depicted in Figure 2. Among the different types of tokens, Security tokens have been utilized for voting rights, patents, copyrights, etc., and tokenized securities for debts, bonds, stocks, and securities. Utility tokens have been utilized for Filecoin, SiaCoin, Golem network, etc., and Currency tokens have been widely deployed to represent fungible and non-fungible assets [9]. The tokens presented by BC tokenization are algorithms implemented as a smart contract on a blockchain. These NFTs represent the ownership of physical or digital assets, such as physical property, virtual collectibles, or negative value assets. Although the NFTs have been defined under the category of currency tokens (shown in Figure 2), these crypto tokens can be used apart for specified purposes such as Multi Token Standard (ERC-1155) [10]. It allows combining fungible and non-fungible tokens in the same token or standards that support royalty

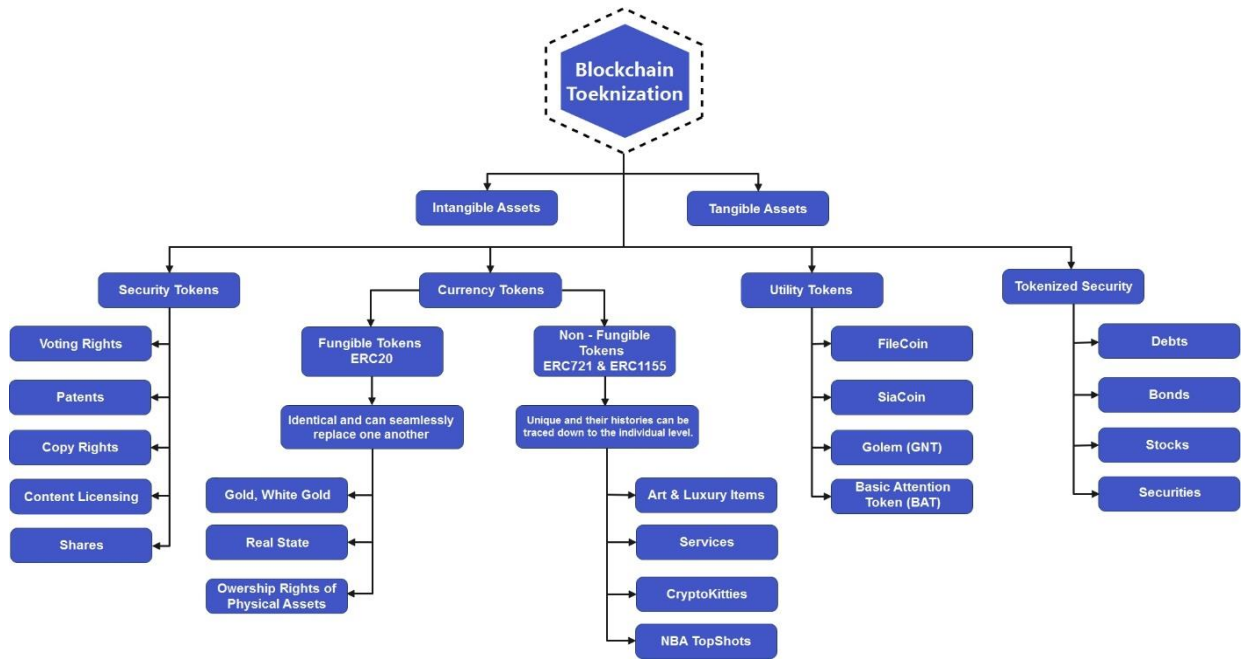

Figure 2. Blockchain tokenization

payments (EIP-2981) [11] and mortgage/rental functions (EIP-2615) [12].

### 1.3 NFTs for Assets Digitization

Figure 3 depicts the generalized architecture of an NFT architecture considering the NFT's well-known project of CryptoPunks. It consists of two roles, i.e., NFT owner and NFT buyer. To digitize an NFT, the owner checks the file, title, and description accuracy. If the correct details are found, the raw data is digitized into a proper format through ERC721 standard-defined functions in the smart contract [13].

The ERC721 functions defined in the NFT smart contract process the creator/owner's request, which stores the raw data in a database external to the BC. However, the owner can also store the raw data in the internal blockchain database, which would be a gas-consuming operation. Once the raw data is stored in the internal blockchain database, the owner signs the transaction, including the NFT data hash. It is then sent to the smart contract, stored in the NFT registry, as depicted in Figure 3. Since NFTs are developed and deployed on BC, thus the consensus layer with logic and blockchain layer for verification is of much importance in the NFT architecture.

At this point, the smart contract from the NFT registry receives the NFT data transaction. It is ready for the minting and trading process. Here logic in the form of transactions is processed to the consensus nodes in a P2P network to attain consensus for privacy. Once verified, the transaction is posted to the synchronizer nodes, which finally post it to the ledger. Once the logic of the ERC721 Token Standard triggers, the NFT data is minted. The transaction confirmation confirms the minting process, which can be traced at any time with a unique blockchain address providing traceability of the digital assets on BC. The ledger provides the traceability of NFTs, which provides a physically defined "digital fingerprint" as a unique identifier. The NFT buyers can transfer the proof of ownership after an approved agreement with the NFT Creator.

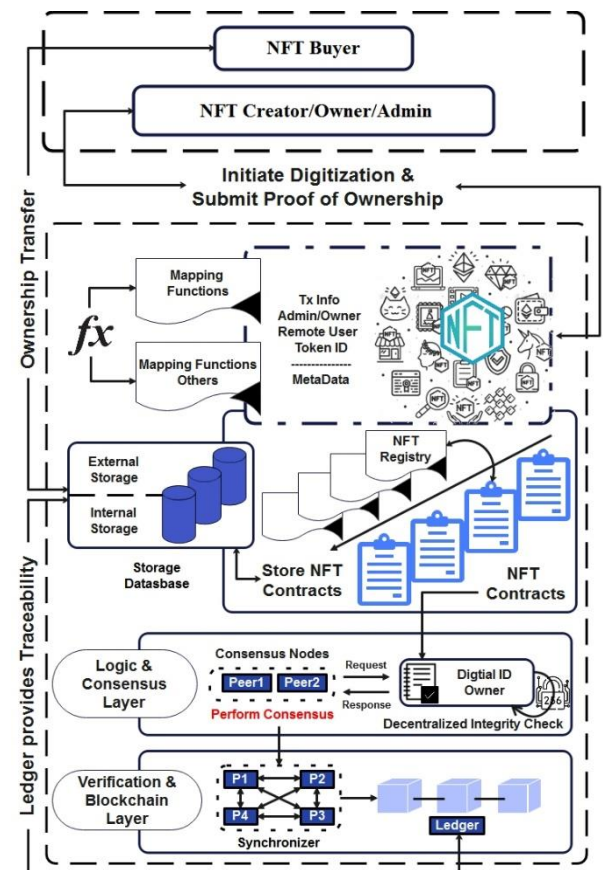

Figure 3. Generalized NFTs architecture

### 1.4 Contributions

In light of the details mentioned earlier for fog computing, blockchain, and blockchain tokenization, the user and device authentication schemes based on decentralized architectures provides a new dimension yet provoke new challenges. The research focuses on the below-mentioned points, which set the major contributions of the article as presented below.
- We propose a novel blockchain-based authentication architecture named Decentralized Smart City of Things (DSCoT) for Owner, Users, fog, and IoT nodes in smart city infrastructure.
- We develop a mechanism of IoT-enabled smart device integration using tokenization in decentralized IoT infrastructure without centralized third-party intervention using blockchain nodes and smart contracts.
    o We utilize NFTs using Externally Owned Addresses (EOA) in blockchain architecture for a digital representation of smart devices.
    o We define additional and newly developed attributes to generate IoT-based NFTs for smart device representation.
    o We introduced mapping to bind the newly developed NFT attributes of Users, fog nodes, and edge nodes with the EOAs of respective devices.
- We devised an authentication *mintNFT()* function, which generates the user NFT (User$_{NFT}$) to represent an authentication access token for the user to access the devices and for the authentication process every time user accesses the nodes assigned.
- Finally, the proposed architecture focuses on the digital representation and authentication of IoT-enabled smart devices from a software standpoint which does not require additional hardware upgrades from the manufacturer, i.e., Physical Unclonable Functions (*PUF*).
    o The proposed architecture actively devises fog computing for all the smart devices connected to a fog node in a particular management system in a smart city, i.e., smart house or smart hospital, etc.

The rest of the paper is organized as follows. Section 2 discusses the literature review of blockchain-based authentication mechanisms with security services and associated problems in smart city architecture. Section 3 presents the novel DSCoT architecture and working methodology of the proposed NFT-based architecture for the user, fog, and smart devices authentication over Hyper Ledger Besu and related architectures. Section 4 presents the implementation and validation of the proposed DSCoT architecture for IoT-enabled smart devices in smart city architecture with results and proof of concept. Finally, a concise conclusion is presented at the end.

## 2 LITERATURE SURVEY

The literature review has focused on blockchain-based authentication schemes and the digital representation of IoT-enabled smart devices. A comparison has been provided to discuss the security challenges these architectures pose regarding IoT-enabled smart assets in a smart city context.

### 2.1 Blockchain-based Authentication Mechanisms

An authentication and access control mechanism based on a distributed architecture for lightweight IoT devices has been proposed in [14]. The mechanism leverages the benefits of fog computing and public blockchain technologies. The proposed mechanism provides a device-to-device (D2D) communication phase for device communication in and out of the system and access control for IoT devices. The Elliptic Curve Digital Signature Algorithm (ECDSA) has been used for key generation, generating public and private keys for the devices and the fog nodes. The security requirements have been tested with the proposed mechanism: Confidentiality, Integrity, Identification, Non-Repudiation, Authentication, and Mutual Authentication.

A proposed framework in [15] *BCoT Sentry* (Blockchain of Things Sentry) integrates blockchain with an IoT network. It enhances network security by analyzing network traffic flow patterns of the device obtained from data storage in the blockchain. The framework has been proposed to keep the lightweight feature of IoT devices which commonly fails to meet computationally intensive requirements for blockchain-based security models. BCoT Gateways are blockchain nodes where an IoT device security module is employed and managed via a smart contract. The authors present a novel approach to the feature selection method (similar feature selection method in machine learning utilizing the maximum information coefficient (MIC), used to measure the discrimination of IoT devices). It captures the IoT device traffic from the network layer and sends this traffic flow feature to the smart contract via blockchain transaction. The smart contract defines the device identification information and related operations and is triggered once the transactions in the blockchain are posted.

A blockchain-based decentralized authentication modeling architecture named *BlockAuth* has been proposed in [16]. The edge devices in the edge layer have been regarded as nodes to form a blockchain network. The authentication scheme claims are suitable for password-based, certificate-based, biotechnology-based, and token-based authentication for high-level security requirement systems in edge and IoT environments. The architecture has been developed using blockchain consensus and smart contract capability.

An Ethereum-based smart contract for edge computing has been proposed as *SmartEdge* in [17] for its low-cost, low-overhead tool for compute-resource management. The authors show the design breakdown of a smart contract into three key steps and describe them in the context of *SmartEdge* design. At the same time, the device representation has been carried out traditionally. The performance was evaluated in terms of low-overhead delay in executing a job and transaction cost in terms of costs that should not be significant relative to their value.

The authors in [18] proposed a new Distributed Anonymous Multi-Factor Authentication (*DAMFA*) architecture that works alternative to SSO (Single Sign-On) and uses a public blockchain (i.e., Bitcoin & Namecoin). The underlying consensus mechanism improves usability, which builds on a Threshold Oblivious Pseudorandom Function (*TOPRF*) for resistance to offline attacks. They claim to include a distributed transaction ledger technology such as blockchain to improve usability. It requires no interaction with the identity provider; hence, the user's authentication no longer depends on a trusted third party. Namecoin blockchain is a public ledger blockchain that allows registering names and storing related values in the blockchain, a secure distributed shared database. A framework for the authentication mechanism based on blockchain has been proposed in [19] named *BCTrust*. It has been designed especially for devices with resource constraints such as computational, storage, and energy consumption constraints. Public blockchain Ethereum has been used to deploy the mechanism to implement the framework. The robustness claimed by the authors is because of the underlying framework of the public blockchain, distributed ledger technology with no central authority for signing the contracts known as smart contracts. These smart contracts provide access control over system identity (SID) authentication mechanisms and User or Device identification (UID).

Blockchain-enabled fog nodes for user authentication schemes have been proposed in [20], which deploy smart contracts to authenticate users to access IoT devices. It is also used to maintain, register, and manage IoT devices, fog nodes, admins, and end-users. The fog nodes provide scalability to the system by relieving the IoT devices from carrying out heavy computation involving tasks related to authentication and communicating with the public blockchain. A distributed system based on the public blockchain design has been proposed with its implementation using Ethereum smart contracts for IoT device authentication at scale. The proposed Ethereum smart contract implements the authentication functionality for adding end-users and IoT assets with the help of an admin who takes care of the overall functionalities and operations of the authentication mechanism.

A blockchain-based decentralized network trust and IoT authentication architecture under the public key encryption system has been proposed [29]. The authors developed the Web of Things (WoT) model that leverages web technologies to improve interoperability and transparency and reduce the chain of trust. A scalable, decentralized IoT-centric PKI has been proposed by combining it with the web-3 authentication and authorization framework for IoT-enabled smart devices.

Table 1. Comparison of Blockchain-based State-of-the-Art Authentication Mechanisms

| Proposed Mechanism | Blockchain Platform | Con. Mech | M/ Auth | Access Control | Data Integrity | Data Anonymity |
|---|---|---|---|---|---|---|
| Blockchain-based Authentication System, 2020 [14] | Ethereum | PoW | ✓ | ✓ | ✓ | ✗ |
| BCoT Sentry, 2021 [15] | Ethereum | PoW | ✗ | ✗ | ✓ | ✓ |
| BlockAuth, 2021 [16] | Hyperledger Fabric 1.4 | PBFT | ✓ | ✗ | ✓ | ✗ |
| SmartEdge- Ethereum, 2018 [17] | Ethereum | PoW | ✓ | ✗ | ✓ | ✗ |
| DAMFA, 2020 [18] | Namecoin | PoW | ✓ | ✗ | ✓ | ✓ |
| BCTrust, 2018 [19] | Ethereum | PoW | ✓ | ✓ | ✓ | ✗ |
| Blockchain-based User Authentication, 2018 [20] | Ethereum | PoW | ✓ | ✗ | ✓ | ✗ |
| WOT, 2017 [21] | Ethereum | PoW | ✓ | ✗ | ✓ | ✗ |
| Blockchain-Based IoT Authentication, 2021 [22] | Ethereum Hyperledger Fabric | PoW / PBFT | ✓ | ✗ | ✓ | ✗ |
| Secure Combination of NFT-PUF, 2021 [23] | Ethereum | PoW | ✓ | ✗ | ✓ | ✗ |

## 2.2 IoT-enabled Smart Device Representation

IoT assets embedded with physically integrated chips (ICs) have been utilized to represent smart devices to mitigate the exploitation of smart assets from physical attacks. This property helps the devices increase security by eliminating the device's physical abuse in the case of adversaries (such as impersonation attacks and side-channel attacks). However, the security comes with a high communication and latency overhead, which limits the use of IoT-enabled smart devices in the CPSs, especially in time-critical applications.

### 2.2.1 Smart Device Representation using Distributed Architectures

Authors in [22] have proposed a blockchain-based platform solution for IoT device authentication, data privacy, and security service via blockchain-based smart contracts. The implementation in Hyperledger Fabric for IC traceability achieved a throughput of 35 transactions per second (TPS). The proposed mechanism uses a defined function on the integrated chips (ICs) named physically unclonable functions (*PUFs*), which imply the authentication mechanism factors. The IoT device hardware was tailored to meet blockchain performance. The authors ensured the smart contract-controlled trust base that the users have private access to their IoT devices and data. A remote configuration of IC features via smart contracts has been devised, where an IC can be programmed repeatedly and securely.

Non-fungible tokens (NFTs) in [23] have been utilized to represent assets by a unique identifier as a possession of an owner. The authors proposed a smart

NFT that is physically bound to its IoT device. This mechanism also defines authentication mechanisms based on Physical Unclonable Functions (*PUFs*), which describe the physical properties of the devices and are used to identify and represent the devices using their private key and BCA address. They have a blockchain account (BCA) address to participate actively in blockchain transactions. These NFTs can establish secure communication channels with owners and users and operate dynamically with several modes associated with their token states. The authors demonstrated the proposal developed with ESP32-based IoT devices and presented the Ethereum blockchain, using the SRAM of the ESP32 microcontroller as the PUF.

### 2.3 Problems Associated with the Smart Device Authentication Mechanisms

The literature survey presents the current state-of-the-art security authentication mechanism for IoT-enabled assets in a distributed IoT architecture. Table 1 depicts an evaluation summary of the proposed authentication schemes.

A. Smart contracts (SC) define applications that are decentralized in nature and are special entities that provide real-world data in a trusted manner. The validation process of these smart contracts could be compromised since the IoT-enabled smart devices can be unstable.
   - SCs in proposed solutions are not designed considering the heterogeneity and constraints present in the IoT-enabled smart devices in the smart city concept.
   - Functions and events in the SCs enable the actuation mechanisms to be employed in the IoT-enabled smart devices much faster.
   - Smart contract deployment with defined authentication functions may provide security, so authentication schemes with smart contacts/decentralized apps (dApps) should be considered.

B. The IoT-enabled smart devices have security issues from the manufacturer's perspective as the asset's firmware is not fully equipped with a security mechanism by default.
   - Especially authentication, access control schemes, and firmware updates are commonly found unattended, posing these assets' exploitation.
   - New strong and lightweight encryption schemes such as SHAIII would help mitigate the authentication and access control issues based on communication and computational costs.

C. Most proposed mechanisms have been deployed on the Ethereum platform, utilizing the traditional Proof of Work (PoW) consensus mechanism. Ethereum is undoubtedly a platform that supports public, private, and hybrid blockchains to be developed and deployed; it also provides the option to utilize decentralized applications (dApps) to provide logic to execute the functions as required. However, the consensus mechanism poses performance issues of fault tolerance, decentralization, stability, and high-level security [24], [25]. Other platforms, such as Hyperledger Besu [26], Hyperledger Fabric [27], Solana [28], etc., provide much more efficient consensus mechanisms for developing solutions over smart contracts.
   - These platforms support more energy-efficient and low latent consensus mechanisms such as IBFT, IBFT 2.0, and Clique.
   - These consensus mechanisms must imply the robust fault tolerance, decentralization, stability, and high-level security and authentication stability of IoT-enabled smart devices to support the smart city infrastructure.
   - The issues with those schemes have also been evaluated based on the security services for collaborative authentication, decentralization, and stability, which depicts most issues relating to access control and data anonymity.
   - These recently proposed mechanisms employ blockchain to attain decentralization but lack robust security and reliability. Hence, there is a need to implement a robust yet reliable consensus mechanism to address blockchain security issues.

D. Recently, Physically Unclonable Functions (*PUFs*), as discussed in Section 2.2, have been the choice to identify devices for solutions implemented on the blockchain.
   - Since PUFs result from hardware modification, it comes with the cost of modifying the device properties and adding manufacturing costs to the budgets, making it hard to develop to implement in smart cities scenarios.
   - As there will be billions of devices connected to the internet, so in the case of smart cities, the manufacturing costs to develop PUFs would not be suitable for Governments and businesses to consider such implementation.

E. In a recent study [23], non-fungible tokens (NFTs) have been utilized to represent assets by a unique identifier as a possession of an owner, but these tokens were employed to bound the IoT assets physically employing PUFs.
   - The assets representation also defines authentication mechanisms based on *PUF*, which describe the physical properties of the devices and are used to identify and represent the devices using their private key and BCA address.
   - The mechanism, however, has not been designed to cater to a complete set of security services (CIA & AAA).
   - The proposed mechanism depends on additional hardware upgrades from the manufacturer, i.e., Physical Unclonable Functions (*PUF*).
   - Since the ERC721 proposed in [23] is hardware-dependent, it requires a hardware

upgrade from the manufacturer, which may incur manufacturing costs.
- The binding of the NFT with the hardware properties may fail the overall system in case of the device malfunctioning.
- With hardware upgrades, the IoT assets have been noticed to have increased initialization time, which incurs latency issues such as initializing Bootloader, located in the main SoC's internal one-time programmable (OTP) memory.
- The coding of the Bootloader cannot be modified since it is the device's Root of Trust (RoT).
- Although the on-chip SRAM, also considered an SRAM PUF, cannot be altered, it still poses time complexity, computational complexity, and latency issues of great concern.

## 3 PROPOSED DSCoT SMART CITY ARCHITECTURE FOR IoT-ENABLED SMART DEVICE AUTHENTICATION

Since the blockchain-based architectures for the representation of admin, users, edge, and fog devices utilizing non-fungible tokens (*NFTs*) in the literature are explicitly lacking, the proposed DSCoT (Decentralized Smart City of Things) utilizes newly defined attributes for representation from a software standpoint omitting the need to update the customer premises equipment (CPE) hardware. The resource constraint nature of the edge nodes (i.e., low processing power, low data storage capabilities, low computational resources, etc.) concerning the digital representation, implementation, and authentication aspect of smart IoT assets in a distributed architecture has been explored.

The NFT functionality has been employed based on the ERC721 standard for smart assets. Despite the development of major categories, the extensions of ERC721 tokens do not define any of the attributes for the smart city infrastructure where users and devices can be identified by a public key and transact uniquely by the identified tokens. The smart contract provides a function-based interface to build non-fungible tokens (NFTs) on the Ethereum blockchain. According to the set objectives, smart city infrastructure based on the distributed network must be explored to provide security for nodes at the sensing and application layers, as depicted in Figure 4. A smart contract has been developed with the functionality of non-fungible tokens (NFTs) for digitally declaring the assets (IoT devices) through Externally Owned Addresses (EOAs) and an authentication mechanism using the SHA-III family encryption protocol. The smart contracts will be deployed to interact with the resource constraint IoT devices based on the decentralized application (dApp) by newly defined NFT attributes for device representation and authentication of devices at the edge of the sensing layer.

Another important mapping aspect has been devised to map newly defined NFT attributes of users, fog nodes, and edge nodes with respective users and devices. It would help attain the security services for authenticating users with fog and edge devices in smart city architecture, i.e., confidentiality, integrity, authorization, and availability. From a software standpoint, the proposed architecture has focused on the digital representation and authentication of IoT-enabled smart devices, which are dependent on additional hardware upgrades from the manufacturer, as seen in the case of [22], [23], i.e., Physical Unclonable Functions (*PUF*).

On the other hand, the platforms based on Ethereum offer much more functions and logic for business models apart from cryptocurrency through smart contract implementation. One main concern of the BC layer is to provide security services, i.e., confidentiality, integrity, availability and authentication, authorization, and audit (CIA & AAA) to the users and CPE (i.e., sensors and actuators). Also, to identify the CPE within CPSs in smart cities in a decentralized manner. The creation of layer-two platforms triggered the invention of the platforms whose architecture is supported and deployed on the Ethereum platform, such as Hyper Ledger Besu (HLB) [26] and Hyper Ledger Fabric (HLF) [27].

In this case, the proposed DSCoT architecture was deployed and tested on HLB because of its two-layered platform capability and the client architecture comprising three major parts: Storage, Ethereum Core, and Networking. Although all components depend on each other, the architecture mainly depends on the client's Ethereum Core. It comprises Ethereum Virtual Machine (EVM) and the consensus mechanisms [29]. HLB implements robust consensus mechanisms such as the Clique, IBFT 2.0, QBFT, and Proof of Authority (PoA) consensus protocols. The second component of the architecture is storage, which exploits integrating a Rock DB key-value database that keeps the data saved on the chain. The third component of the architecture is networking which supports peer-to-peer (P2P) communication with other nodes utilizing the devp2p protocol. It helps in client-to-client communication in BC. The stored private transactions on Besu through Tessera provide a private transaction manager to implement privacy.

The Besu network was chosen for the proposed architecture as its important from a privacy perspective where each node sending or receiving private transactions requires an associated Tessera node [30]. Deploying a personal blockchain is like a real blockchain connected to a public blockchain, which runs on the computer in the closed network and provides connectivity that runs on the machine. Deploying HLB was helpful as it is a valuable tool for testing and implementing blockchain-based decentralized apps.

### 3.1 Working of Proposed DSCoT Architecture

The working methodology is innovative since a novel NFT standard has been presented in DSCoT architecture (Decentralized Smart City of Things) for the smart city. The DSCoT proposes a standard for deploying smart device representation through the proposed non-fungible tokens mechanism utilizing NFT-based EOAs and their authentication mechanism via smart contracts. The implementation of a distributed technology blockchain has been carried out, as presented in Figure 4. The decentralized application (dApp) functions at the application layer by deploying smart contracts stored in the NFT registry in blockchain storage and responding as required by respective CPSs, such as smart hospitals, smart homes, smart industries, smart cars, etc. It also ensures functionality by deploying smart contracts to authenticate the owner and remote users at the application layer and authenticating the fog and customer premises equipment (CPE) at the sensing layer, as shown in Figure 4. The blockchain-enabled smart city architecture can be classified into four layers while including the blockchain layer supports robust security mechanisms, as discussed in Section 2.3. The blockchain-based architecture presented in the figure adds a BC layer to the generalized smart city-layered architecture. It integrates IoT-enabled smart devices in blockchain-enabled CPSs (such as smart homes, hospitals, etc.). As shown, the sensing layers deploy the edge and fog nodes (i.e., sensors, aggregators, and actuators) in the physical environment within cyberspace that supports actuation based on the data collection. Here fog computing provides enough computational resources for data collection and processing for environmental sensing. The network layer provides connectivity using communication and transmission technologies at the transmission layer, while the command and control work on the application layer defines the applications for the asset's behavior at the physical layer. The proposed architecture provides security in terms of confidentiality, data integrity, availability, and anonymity using the SHA-III family one-way encryption mechanism, as discussed in Sections 3.3.1 and 4.1. The blockchain layer deploys the Hyperledger Besu, enabling the proposed architecture's distributed functionality. NFT-based EOAs in this layer represent these components which are unique and non-interchangeable units of data stored on a distributed ledger. Thus, utilizing the proposed blockchain tokenization in DSCoT infrastructure without a centralized third-party intervention provides a mechanism to digitally define the assets and components to attain a robust authentication mechanism.

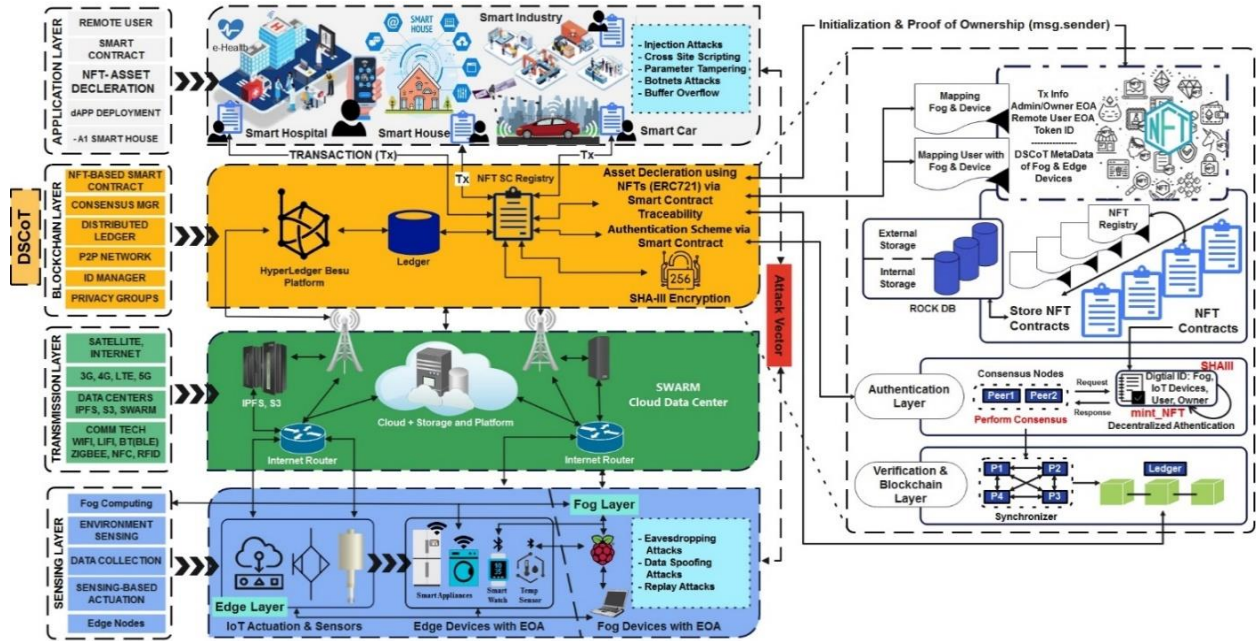

Figure 4. The proposed Decentralized Smart City of Things (DSCoT) architecture

### 3.2 Proposed DSCoT NFT Mechanism

Figure 4 shows the NFT functionality, which depicts asset initialization, digitization, and authentication mechanism with users. As shown in the figure, the assets initialization triggers if the proof of ownership is verified (msg.sender). The owner is the creator/admin of the smart contract who can add, delete or map the fog devices with the IoT assets. The transaction (Tx) info, details of NFT-based EOAs of the User, fog, IoT assets, DSCoT metadata, and $Token_{ID}$ are stored in the NFT registry to authenticate the assets accordingly. The internal storage is utilized to store the NFT registry. In contrast, the external storage is utilized to access the NFT registry externally by the users, as demonstrated in Figure 3 and Figure 4, respectively.

The authentication layer shown in Figure 4 adds the authentication and authorization mechanisms that deploy a decentralized application to provide the authentication logic for the connected nodes in CPSs in a smart city context. Specifically, the proposed architecture will help integrate robust authentication of CPE, exploiting the SHAIII encryption protocol functionality. The encryption protocol has been deployed in the mintNFT function at the blockchain layer in DSCoT. Its additional uses for the function, such as an authenticated encryption system, would leverage faster hashing in the proposed architecture. Since centralized systems, including key management systems, may jeopardize the system's security because of trusted third-party service providers, the cryptosystems based on decentralized technology have opted to enable the solutions deployed on top blockchain solutions.

Once user authentication utilizing the mintNFT function with fog and IoT assets completes, the consensus mechanism triggers, which in this case is IBFT 2.0. The transaction is posted to the peers in the P2P network, where synchronizer nodes synchronize it with its group before it is posted to the blockchain ledger as an immutable transaction. The posted transactions would provide traceability as unique identifying codes are the property of an NFT. It enables every asset's digitization and can be traced in the ledger.

### 3.3 Design & Implementation of Proposed DSCoT

Smart contracts help develop a client-side application that runs on top of the blockchain as a decentralized app (dApp). These applications are developed in a solidity programming language that acts more like Java Script. Remix IDE (v0.23.3) has been used to develop, compile and deploy the proposed mechanism via smart contracts as an ERC721-IoT standard. As discussed in Section 1.2, non-fungible tokens, or NFTs, are built on the Ethereum Request for Comment ERC-721, which defines a standard interface using wallet applications to work with any NFT on Ethereum platforms. ERC-721, in contrast to its predecessor, the ERC-20 (fungible and interchangeable) tokens, are non-interchangeable and have uniqueness for each assigned asset. This lucrative property makes its use for smart devices distinction (non-fungible).

The proposed DSCoT represents the smart devices with $token_{id}$ and NFT-based Externally Owned Addresses (EOA) referred to as resource owners. NFTs provide two basic attributes for the identity and uniqueness of assets, i.e., token identification ($token_{id}$) and EOAs that can be owned, transferred, and approved to act on their behalf. However, additional attributes represent the owner, remote users, and the devices at fog and edge layers (IoT-enabled smart devices), with functions defining the functionality, are explicitly absent. Hence, a novel NFT-based architecture is proposed for all assets; also, more attributes have been devised that would be helpful for asset representation and the authentication mechanism to validate the device's authenticity, as depicted in Table 2.

Table 2. DSCoT proposed metadata

| Sr # | Attribute | Description |
|---|---|---|
| 1 | Owner | EOA of the Owner |
| 2 | $token_{Id}$ | Token ID of the Owner |
| 3 | $U_{ID}$ | User Identification |
| 4 | $D_{ID}$ | Smart Device Identification |
| 5 | $Fog_{ID}$ | Fog Device Identification |
| 6 | T | Time Stamp |
| 7 | ΔT | Change in Time duration |

The DSCOT metadata of the standard attributes defined in the proposed ERC721 standard, such as Owner (address) and token ID, are utilized in DSCoT to validate the owner for managing the resources. The $U_{ID}$ attribute was created to represent user identification, and $D_{ID}$ represents device identification. Similarly, $Fog_{ID}$ represents the fog node identification while T and ΔT represent the block timestamp and change in time for the blockstamp to record the replay or spoofing attacks. The attributes' descriptions have been defined and shown in Table 2 for better understanding.

#### 3.3.1 Components and Mechanisms of Proposed DSCoT Architecture

The smart contract (SC) of the proposed DSCoT has been designed to expand its functionality to different CPSs in smart city architecture. Hence the designed components can be integrated as required, such as in smart homes, smart hospitals, smart supply chains, smart industries, smart cars, etc.

The main components are the proposed smart contract, the owner (admin), the user, the fog device, and the IoT-enabled smart device. As depicted in Table 3, the components with functions and events are defined in the interfaces, while the main functions were developed in DSCoT SC. The ERC721 was imported using the OpenZeppelin Contracts, which provide flexibility regarding combining these as useful custom extensions [31]. The components and mechanisms of the proposed DSCoT have been presented with algorithms and pseudo-codes for clear understanding.

The pseudo-code steps shown in Algorithm 1 depict the initialization of the DSCoT parameters and definitions of the components during the deployment of the SC. It shows the initialization of the Resource Owner/Admin, the only entity that can initiate the smart contract approved with initial DSCoT *approve* operators, i.e., token identification ($token_{id}$) with an externally owned account (EOA). It is defined in the devised constructor to ensure the contract's confidentiality, availability, and authorization to own and execute by this ID or otherwise reject the initialization. Once initiated, only the owner can update/add/delete and call the functions. Furthermore, lists and structs for admins, tokens, devices, and mapping functions have been devised for asset representation in the proposed mechanism.

---

**Algorithm 1:** *Initialization of DSCoT Params and Components Definition*

**Params:**

**constructor: admin / owner == msg.sender**
*// creator of contract as the first admin/owner*

**admins [ ]:** *// admins of the system*

**struct Token { }:** *// struct for the information of a given token*

**struct Devices { }:** *// struct for the addresses of devices*

**Token[ ]:** *//list of all the issued tokens*

*// mapping for Users and their accessible devices*
**mapping (address => Devices[ ]) user_devices;**

*// mapping for devices at a fog node*
**mapping (address => address[ ]) fog_devices;**

**modifier onlyOwner { }:** *// for user check at modifications*
**bool** admin = false;
**Loop** through admins.length;
    **If** (msg.sender == admins[i])
       admin = true;
       **BREAK the Loop;**

```
If (!admin)
    revert("Not an Admin");

else (end);
```

The metadata in Table 2 and 3 includes the transaction payload in terms of NFT-based EOAs that employ the list of operations as depicted in the below-mentioned algorithms.

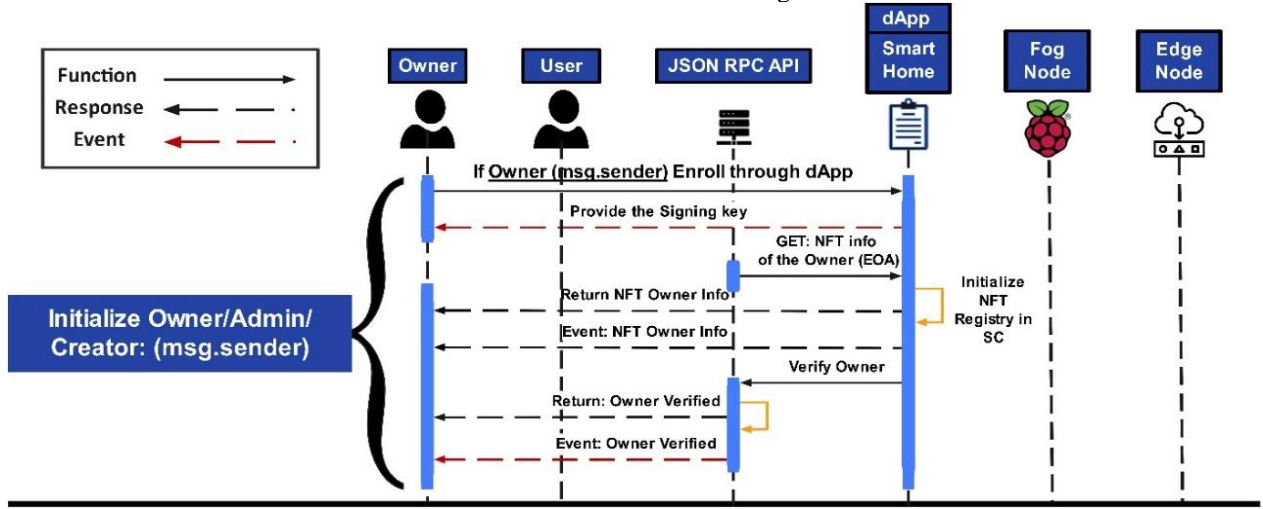

Figure 5. Sequence diagram of session connections for initializing Owner Representation

Depicted in Figure 5 is the process of initializing the owner and the secure session with the SC. The sequence diagram shows that the owner is the only admin or creator of the SC and will only initiate communication sessions for authorized access or will be rejected otherwise. It provides the owner the authorization to access a particular SC for execution in the proposed DSCoT architecture. Once verified, the owner provides the public and private key pair, which would initiate requesting the NFT info to generate the NFT for the owner. Once TokenID is assigned, the information will be saved in the NFT registry, and details will be returned to the owner with an event. As aforementioned, the SC functions would be deployed based on the proposed mechanism of the ERC-721-IoT standard by the resource owner, who will be an admin in this context. Table 3 presents the functions that have been deployed.

Once the owner initializes the smart contract (SC), the assets must be verified. It would initiate the mapping function as a next step to map the verified fog node with verified IoT assets, as depicted in Algorithm 2.

**Algorithm 2:** *//Assigning an IoT Node to Fog Node*
**@DeviceFogMapping(EOA$_{fog}$, EOA$_{device}$)**
Public Virtual override OnlyOwner
fog_devices[ ].push(EOA$_{device}$);
   **emit** FogDeviceMappingAdded(EOA$_{fog}$, EOA$_{device}$, EOA$_{admin}$);

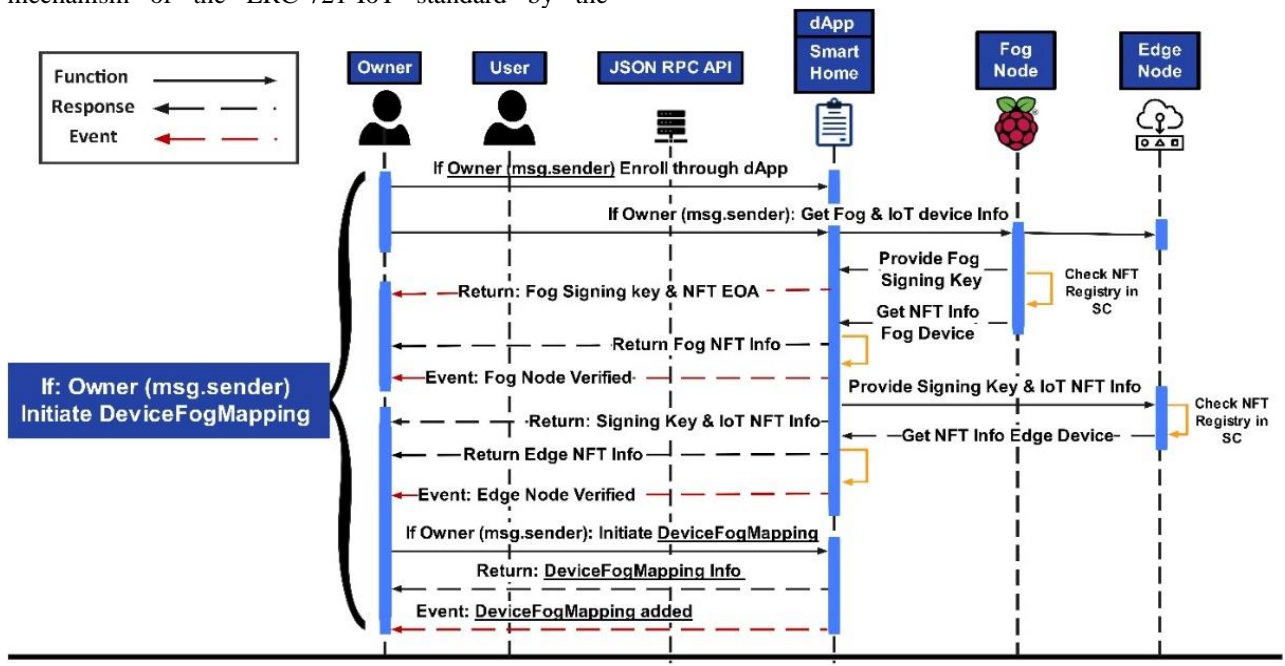

Figure 6. Sequence diagram of session connections for DeviceFogMapping representation by the owner

Events with the NFT EOA parameters ($EOA_{Fog}$, $EOA_{Device}$) will be generated and saved in the NFT registry. The next step is to initiate the IoT asset and fog device mapping, which would be initiated by the *DeviceFogMapping()* function that maps IoT devices with respective fog nodes with the NFT EOAs parameter ($EOA_{Fog}$, $EOA_{Device}$) as depicted in Algorithm 2.

Figure 6 depicts the sequence diagram of the owner initiating a secure session connection to authorize the fog device with the IoT-enabled smart device via mapping. The figure depicts the access authorization process only if the smart contract owner initializes the secure session with the smart contract, or the access will be rejected otherwise. The owner queries the public and private key pair and NFT EOA information of the fog device in the first step. Once verified, the same procedure will be followed for the IoT assets in the second step. The *DeviceFogMapping()* function initiates that maps IoT devices with respective fog nodes with the NFT EOAs parameter ($EOA_{Fog}$, $EOA_{Device}$), and details are returned to the owner with events, as depicted.

The next step is to assign the mapped devices to a user who can access these devices within the respective CPS. The owner gets the verified user ($EOA_{User}$) and DeviceFogMapping ($EOA_{Fog}$, $EOA_{Device}$) information, and if the information is matched, the *UserDeviceMapping()* function will be initialized. This function would map the user with the fog and the IoT-enabled smart device via mapping to provide access to these devices once the authentication phase competes in the next step. At this point, the *UserDeviceMapping()* maps the user with respective fog and the IoT-enabled smart device with the NFT EOA parameters ($EOA_{Fog}$, $EOA_{Device}$), as depicted in Algorithm 3. Events with the NFT EOAs ($EOA_{User}$, $EOA_{Fog}$, $EOA_{Device}$) will be generated and saved in the NFT registry. Algorithm 3 shows the pseudo-code for the mapping process flow of the *UserDeviceFogMapping()* function. The User NFT EOAs parameter ($EOA_{User}$) is assigned to the respective fog node. The fog node must have an IoT asset assigned to assign the user, or the request will be denied otherwise, as depicted in Algorithm 3.

---

**Algorithm 3: //Assigning a User to a Fog Node with an assigned IoT Node**

@**UserDeviceFogMapping**($EOA_{user}$, $EOA_{fog}$, $EOA_{device}$)

Public Virtual override OnlyOwner
**bool** deviceExists = false;
**Loop** through fog_devices[ ].length;

**If** fog_devices[] == device;
  deviceExists == true;

**BREAK** the Loop;

**If** (deviceExists)
  users_devices [ ].push(Devices($EOA_{fog}$, $EOA_{device}$);
  **emit** UserDeviceMappingAdded($EOA_{user}$, $EOA_{fog}$, $EOA_{device}$, $EOA_{admin}$);

**else**
  **emit** DeviceDoesnotExist($EOA_{fog}$, $EOA_{device}$, $EOA_{admin}$);

---

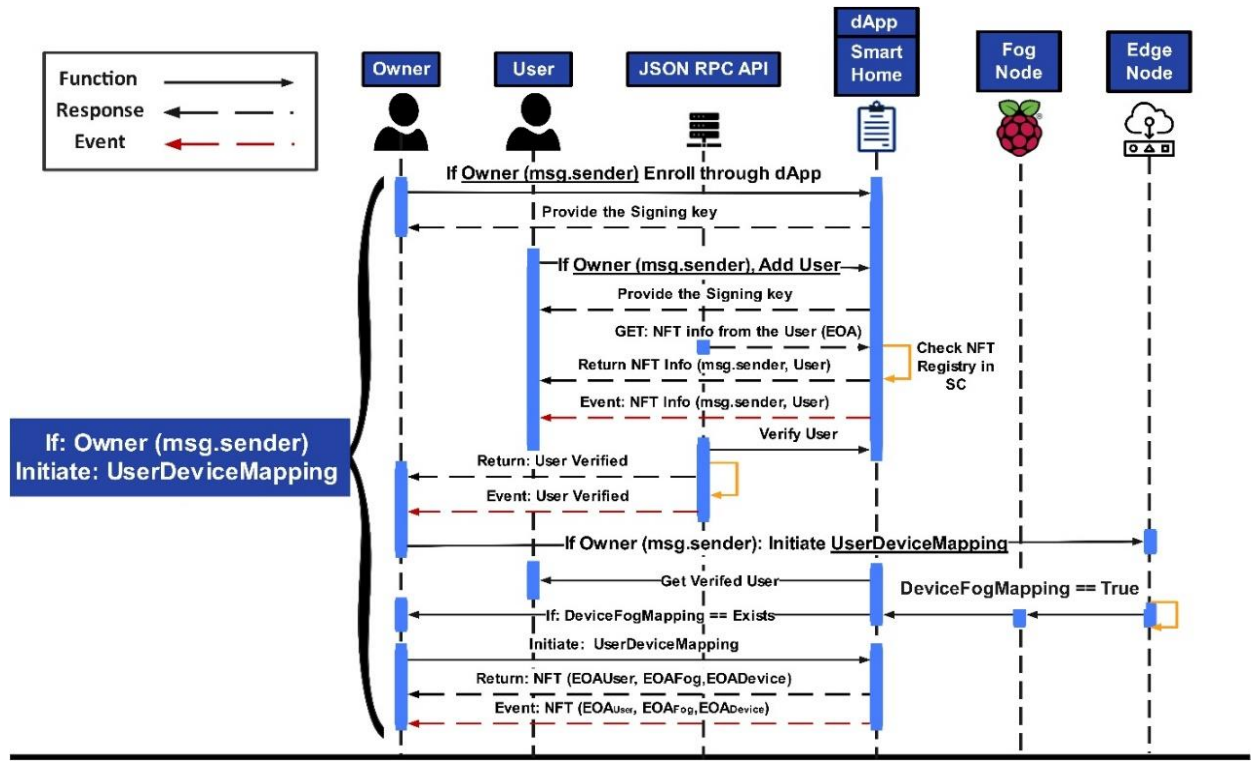

Figure 7. Sequence diagram of session connections for UserDeviceMapping representation

Figure 7 depicts the owner initiating a secure session connection to add the user with its assigned NFT-based EOA and provides the public and private key pair, which, once verified, would initiate to request the NFT info for the user. Once $Token_{ID}$ is assigned, the information will be saved in the NFT registry, and details will be returned to the owner with an event.

Once the IoT-enabled smart devices are mapped with the fog device and the respective user, the NFT minting function must be triggered to authenticate users to access the devices.

Algorithm 4 shows the *mintNFT()* function, which generates the user NFT (User$_{NFT}$) to represent an authentication access token for the user to access the devices and for the authentication process every time user accesses the nodes assigned. It is the final step where the authentication process will trigger by checking the EOA$_{User}$, EOA$_{Fog}$, and EOA$_{Device}$ in the respective lists, as shown in Algorithm 3. An NFT for the user (User$_{NFT}$) will be generated utilizing the SHAIII encryption protocol, which utilizes an authenticated encryption system as presented in Algorithm 4. It will authenticate the assets once the NFT-based EOAs of the users, fog, and IoT devices are mapped with each other or will reject the authentication request otherwise.

The generated NFT$_{Id}$ will be a unique identification access code used for user authentication whenever the user wants to access the devices, as depicted in Algorithm 4.

**Algorithm 4:** *// Mint Function to Create NFTs for UserFogDevice Authentication*
**@mintNFT(EOA$_{device}$, EOA$_{fog}$)**

Public Virtual override OnlyOwner
**bool** deviceExists = false;

**Loop** through fog_devices[].length;
   **If** fog_devices[] == device;
     deviceExists == true;
     **BREAK** the Loop;

  **If** (!deviceExists)
    emit DeviceDoesnotExist(EOA$_{fog}$, EOA$_{device}$, EOA$_{admin}$);
  else
    **bool** auth = false;
    **Loop** through users_devices[EOA$_{admin}$].length;
      **If** users_devicesEOA$_{admin}$][].device == device;
        auth == true;
    **BREAK** the Loop;

**If** (auth) // shares successful authentication event
  bytes32 _tokenID=keccak256(abi.encodePacked
        (EOA$_{device}$, EOA$_{fog}$, EOA$_{admin}$, block.timestamp));
  **emit** Authenticated(EOA$_{admin}$, EOA$_{device}$, EOA$_{fog}$);

  Tokens.push(Token(_tokenID, block.timestamp));
  **emit** TokenCreated
        (tokenID, EOA$_{admin}$, EOA$_{fog}$, EOA$_{device}$, block.timestamp);

**else if**(!auth) // trigger failed authentication event
  **emit** NotAuthenticated(EOA$_{admin}$);

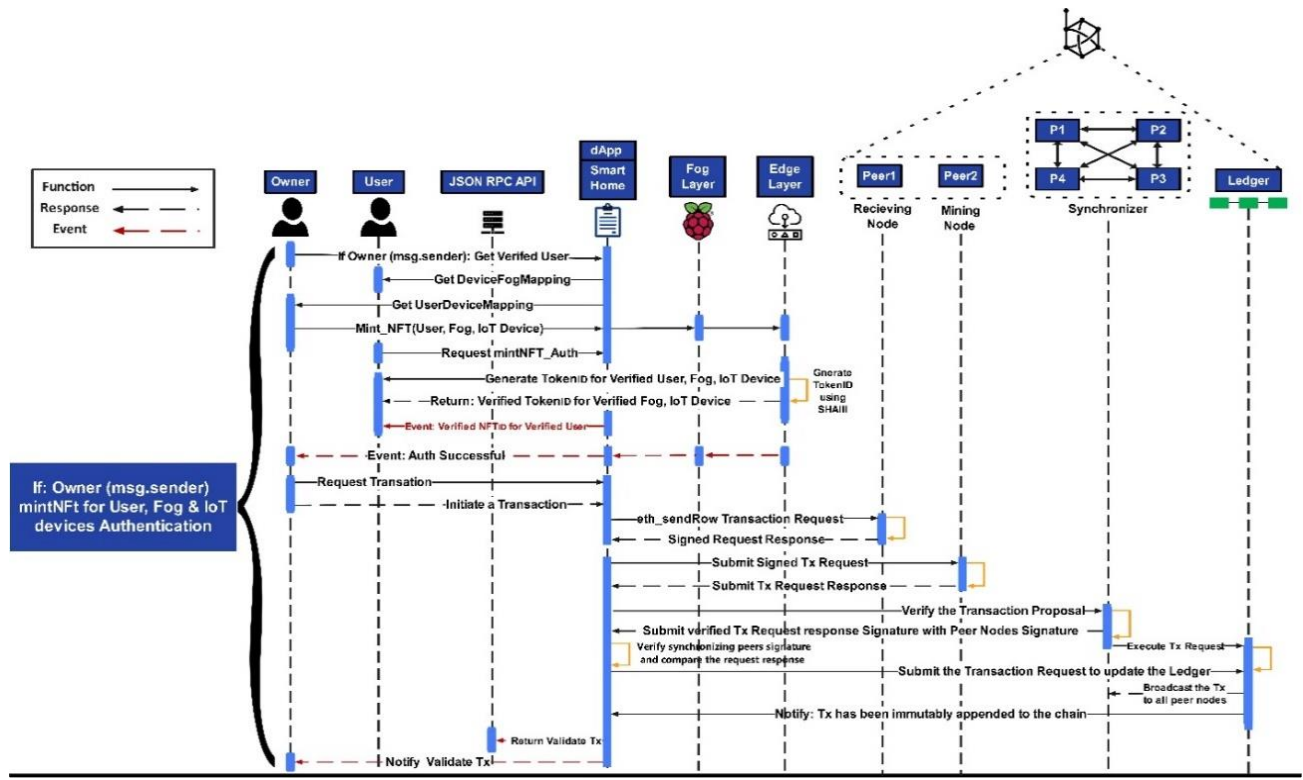

Figure 8. Sequence diagram of session connections for mintNFT_Auth representation

Figure 8 depicts the mapped user initiating a secure session connection to authenticate and generate NFT for user authentication. A complete process of posting transactions has been shown in the sequence diagram, so a complete blockchain working may be presented. The pseudo-code in Algorithm 4 shows the function to verify if the IoT nodes are assigned to the fog nodes and the user. The request to authenticate the user will be denied if no device is detected in the fog list. Upon NFT authentication, the node signs it with the user's account public key (User$_{PK}$).

- NFT$_{pass}$ = Pass$_{User}$(User$_{NFT}$), User$_{NFT}$ whereas, User$_{NFT}$=(UID, T, ΔT, EOA$_{User}$, EOA$_{Device}$, EOA$_{Fog}$, User$_{PK}$).

The user signs the token with the account's private key (User$_{IK}$), as shown below, and the user is authenticated to access the mapped devices.

- $NFT_{pass} = Pass_{User}(User_{NFT})$ whereas, $User_{NFT}$= ($Token_{Id}$, T, ΔT, $EOA_{User}$, $EOA_{Device}$, $EOA_{Fog}$, $User_{IK}$).

The $NFT_{Pass}$ represents the user's non-fungible token generated by incorporating the user's token, block timestamp, and change in block timestamp together with NFT-based EOAs of the users, fog, IoT devices, and the user's EOA private key. It generates an authentication access token for the user to access the devices and for the authentication process, every time user accesses the nodes assigned.

An innovative approach of *call()* methods has been devised to query the smart contract for the status of assets in the NFT registry and offers no transaction cost (in Ether/Gewi). It makes the proposed architecture efficient in terms of time complexity. Only the owner can perform the call operations; otherwise, the request will be rejected. The events will be emitted once a specific operation has been performed. The smart devices (SDs) define the NFT-based EOAs mapping mechanism with a particular user EOA, ensuring security services as discussed in Section 4.1.

Once a successful event has been generated, the owner request to initialize the transaction to append the blockchain, as shown in Figure 8. The transaction request is generated to a p2p network of receiving nodes, which sends the signed request response. The transaction with a signed request-response is submitted to the p2p network of mining nodes which verifies the transaction and submits the transaction proposal to a synchronizer p2p network. It verifies the synchronizing peers' signatures and compares the request response to verify it. Once the response is verified, it is posted to the ledger.

After the transaction record has been appended to the ledger, the updated transaction request response is broadcasted to all the peers to synchronize the transaction and append the chain to attain immutability. The validated response is returned and notified to the owner via an event.

Table 3. Components & functions of the proposed DSCoT architecture

| | |
|---|---|
| *Functions & Events to add/Del and check the No. of Admins NFT EOAs* | function approve(address _approved, uint256 _tokenId) external payable; |
| | event AdminAdded(address indexed newAdmin, address indexed addingAdmin); |
| | event AdminAlreadyExists(address indexed newAdmin, address indexed sender); |
| | function No_ofAdmins() external view returns (uint256); |
| | function adminAdd() external view returns (address[] memory); |
| | function delAdmin (address admin) external; |
| | event AdminDeleted(address indexed newAdmin, address indexed deletingAdmin); |
| *Functions & Events to Add/Del/Map devices (IoT, Fog)* | function DeviceFogMapping(address fog, address device) external; |
| | event FogDeviceMappingAdded(address indexed fog, address indexed device, address indexed addingAdmin); |
| | event FogDeviceAllMappingDeleted(address indexed fog, address indexed deletingAdmin); |
| | event DeviceDoesnotExist(address indexed device, address indexed fog, address indexed sender); |
| | function delDev(address fog) external; |
| *Functions & Events to add/Del/Map Users with Smart devices* | function UserDeviceMapping(address user, address device,address fog) external; |
| | event UserDeviceAllMappingDeleted(address indexed user, address indexed deletingAdmin); |
| | event UserDeviceMappingAdded(address indexed user, address indexed device, address addingAdmin, address indexed fog); |
| | function delUser(address user) external; |
| *Functions & Events to check balance and owner of a token.* | function balanceOf(address _owner) external view returns (uint256); |
| | function ownerOf(uint256 _tokenId) external view returns (address); |
| | function tokens_Issued()public view returns (Token[] memory); |
| *Minting Functions & Events for User and devices Authentication Mechanism* | function mintNFT(address device, address fog) external; |
| | event Authenticated(address indexed user, address indexed device, address indexed fog); |
| | event NotAuthenticated(address indexed user); |
| | event InvalidUser(address indexed device, address indexed fog, address indexed sender); |
| | event TokenCreated(bytes32 indexed _tokenID, address indexed User, address device, address indexed fog, uint256 timestamp); |

## 4 RESULTS AND DISCUSSION

### 4.1 Security Services

Most of the proposals did not address security services such as confidentiality, integrity, and availability (CIA) and are reliant on the default security mechanisms, as shown in Table 1. Some rely only on the basic consensus mechanism, while these proposals achieve integrity by implementing the encryption protocol, as shown in Table 5. As discussed in Sections 3.3 and 3.3.1, the security services (CIA) and authorization for DSCoT architecture have been achieved. Table 4 further presents achieved security services.

### 4.1.1 Confidentiality and Availability

As shown in Algorithm 1 in Section 3.3.1, to achieve the security service, a constructor has been devised to ensure the smart contract's (SC) confidentiality and availability, which would only allow the Creator/Admin/Owner to own the smart contract. The owner will be able to execute the functions by its ID or reject the initialization. Once initiated, only the owner can update/add/delete/call other admins, users, IoT assets, and fog devices. This property helped achieve confidentiality, while the availability of SC and assets restricted only to the owner of the SC does not allow the resource availability to anyone apart from owners.

### 4.1.2 Authorization

Algorithm 1 further depicts a modifier devised and defined as "OnlyOwner." It adds another layer of security as the owner (msg.sender) defined in the contract's constructor will be registered as the only owner who can log into the smart contract. It accesses all the smart contract functions for adding users and smart devices. It ensures only approved (DSCOT *approve()*) admin/onlyOwner as an authorized entity to initiate all the functions requests. Otherwise, the access will be denied, which fulfills the need for the authorized user, such as the owner, to achieve authorization in smart city architecture.

### 4.1.3 Integrity

The proposed architecture has been integrated for robust authentication, exploiting the SHAIII encryption protocol functionality as depicted in Algorithm 4. It has been deployed in the mintNFT function at the blockchain layer, whose additional uses for the function, such as an authenticated encryption system, would leverage faster hashing in the proposed architecture. The process generates the NFT Token$_{Id}$ for the user using the SHA-III algorithm. The generated NFT Token$_{Id}$ will be a unique identification code used for user authentication whenever the user wants to access the devices.

Table 4. Security services provided by DSCoT architecture

| Security Services | Protection |
|---|---|
| Confidentiality | Achieved by devising a constructor using SC |
| Integrity | Implements Encryption Protocol - SHA III |
| Availability | Achieved by devising a constructor using SC |
| Traceability | Achieved using Hyperledger Besu P2P Synchronizer Network |
| Authorization | Achieved by devising a modifier for "OnlyOwner" using SC |

## 4.2 Validation of Proposed DSCoT Architecture

After the testbed was deployed, the methods imposed by NFTs using smart contracts need to be validated by the amount of Gas consumed in carrying out the transactions on the private Besu platform. The Ethereum blockchain platform uses the cryptocurrency ether (ETH), while the smaller fractions are measured in Gwei. Gas is the execution cost of the operation that needs to modify the data on the blockchain. The decentralized app (dApp) execution, such as a smart contract spends Gas to allocate resources defined.

A lightweight decentralized app implementation would cost a lesser Gas limit, which means less work to execute a transaction using ETH (Ether) via smart contract. More Gas would be consumed, resulting in an inefficient solution. DSCoT functions evaluation at the time of deployment was carried out for the gas consumption so that the cost of each function may be known.

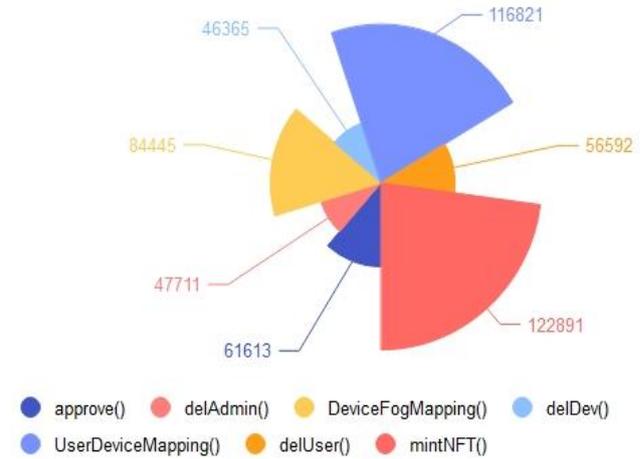

Figure 9. Proposed DSCoT Gas consumption

Figure 9 shows the Gas consumed by the main functions in the proposed DCoT, where the mint function has consumed more Gas which is expected for the encryption and authentication of users and devices. UserDeviceMapping maps the user to the respective fog, and IoT nodes, have also consumed more gas. In contrast, the rest of the functions, such as *approve()*, *delAdmin()*, *delDev()*, *delUser()*, and *DeviceFogMapping()* functions have consumed almost the same amount of Gas on average.

## 4.3 Efficiency Performance

As per the literature review in Section 2, it is evident that the digitization of the IoT-enabled smart assets and authentication mechanisms utilizing NFTs is lacking; hence, comparing the proposed DSCoT would not be possible. However, as presented in Table 5, a solution based on NFT for linking the IoT assets has been proposed in [23], utilizing the hardware modifications in the IoT assets as in the use of physically unclonable functions (PUFs). It is used for IoT asset identification. Since the solution depends on binding the NFT with PUFs to identify an IoT asset and authentication mechanism, the results show high latency in terms of device initialization time. The functions and components in the proposed architecture have been observed to have consumed more gas to perform the transactions to append the data on the blockchain [23].

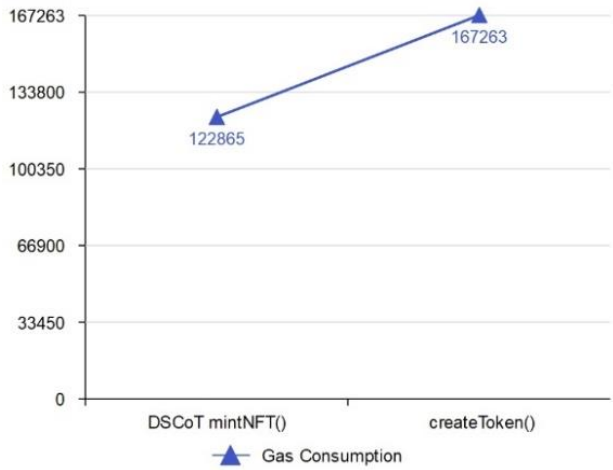

Figure 10. Gas consumption ~ DSCoT mint

A comparison of Gas consumption for the proposed DSCoT in contrast to the PUF-based NFTs has been presented in Figure 10, which depicts a considerably low gas consumption for main minting functions, which evidently shows DSCoT *mintNFT()* function is more efficient as it consumes low Gas (122865) than the PUF-based NFT *createToken()* function, which consumes more Gas (167263) comparatively.

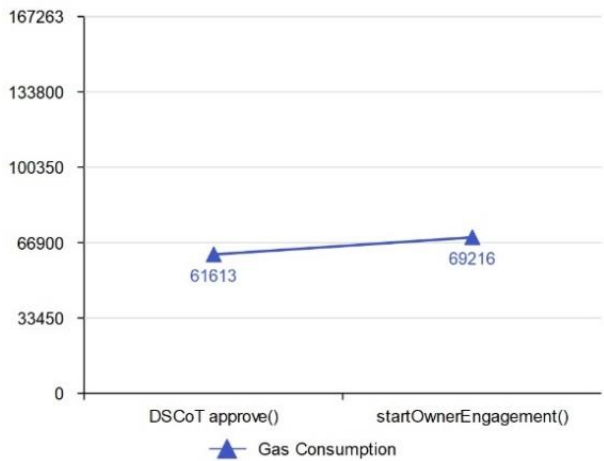

Figure 11. Gas consumption ~ DSCoT approve

DSCoT *approve()* function in Figure 11 also shows considerably low gas consumption (61613) for the proposed DSCoT in contrast to the *startOwnerEngagement()* function, which comparatively consumes more Gas (69216).

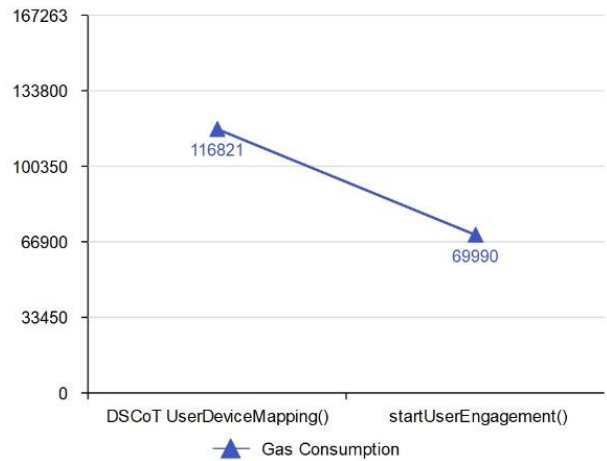

Figure 12. Gas consumption~UserDeviceMapping

The *UserDeviceMapping()* function in Figure 12, on the other hand, consumes low gas consumption (116821) but comparatively depicts more than the PUF-based NFT's (69990) *startUserEngagement()* function. The rise has been observed due to the *User* and *Device* Mapping that takes place at this stage simultaneously in DSCoT, while the *startUserEngagement()* in PUF-based NFTs consumes low gas because it is used to engage the *user* only.

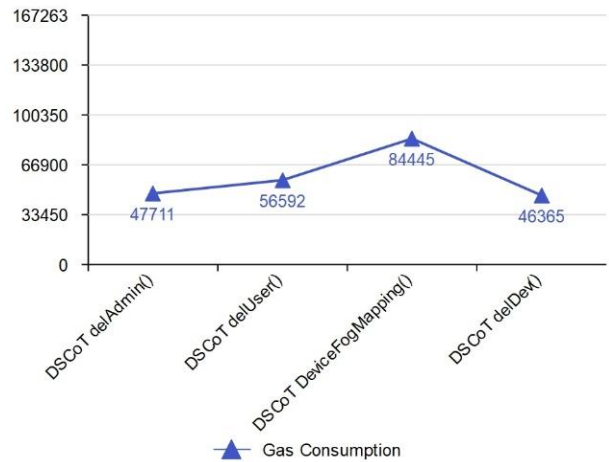

Figure 13. Gas consumption ~ call to NFT registry

Figure 13 shows the gas consumption for proposed DSCoT functions that have been created to check the status of the NFT registry with considerably low gas consumption. These functions could not be compared with other functions of PUF-based NFT solution as they could not be compared in terms of their utilization and functionality; however, the depicted functions in the proposed DSCoT remain efficient in terms of low gas consumption.

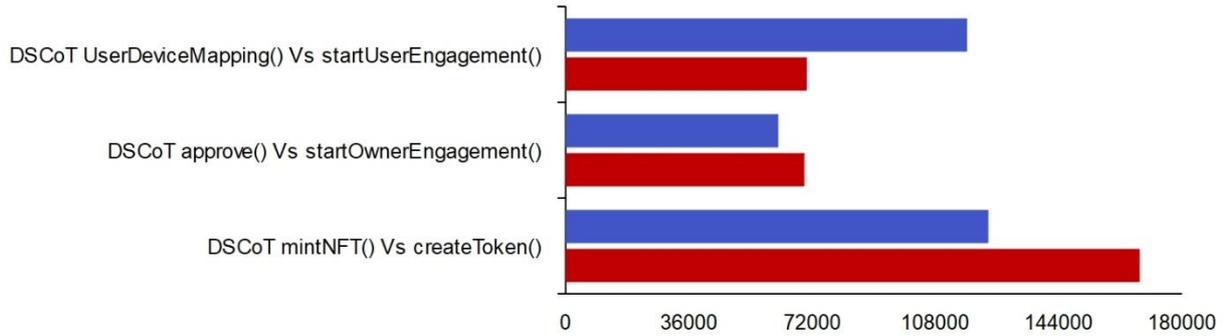

Figure 14. Gas consumption of proposed DSCoT vs. PUF-based NFT functions

Figure 14 provides a better graphical view of the Gas consumption of functions in the proposed DSCoT (*BLUE*) and PUF-based NFT (*RED*), which depicts a considerably low gas consumption for the proposed DSCoT. As shown in the graph, DSCoT *mintNFT()* vs. *createToken()* was observed to be approximately ≈ 27% cost efficient while a DSCoT *approve()* vs. *startOwnerEngagement()* was observed to be approximately ≈ 11% more cost efficient respectively. Furthermore, Table 5 compares state-of-the-art BC-based solutions with the proposed DSCoT, which explicitly implements the Hyperledger Besu to make the NFT-based architecture robust by employing the Istanbul Byzantine Fault Tolerant (IBFT 2.0) consensus mechanism. At the same time, most of the solutions have been deployed on public BC with a default consensus mechanism. It opens doors to performance issues of fault tolerance, decentralization, stability, and high-level security, as discussed in Section 2.3. The private blockchain implementation attains consensus within hundreds of milliseconds, providing low latency. This property of the consensus mechanism is crucial for building blockchain-based IoT networks that provide low communication overheads and fault tolerance [32].

In contrast, among all the BC-based solutions, only the authors in [23] and the proposed DSCoT managed to employ an NFT-based solution for user and IoT-enabled smart device authentication mechanisms. However, the authors in [23] implement the default consensus mechanism of Proof of Work (*POW*) which does not look promising for IoT networks because of the high energy consumption, communication overheads, low fault tolerance, and latency issues [43].

Apart from the proposed DSCoT architecture, none of the solutions manage to deploy all the features of security services (i.e., confidentiality, availability, and integrity), making it more robust in terms of security.

Table 5. DSCoT comparison with related State-of-the-Art blockchain-based mechanisms

| Proposed Mechanism | Blockchain Platform | Consensus Mechanism | NFT | Time Complexity | CIA |
|---|---|---|---|---|---|
| Blockchain-based Authentication System, 2020 [14] | Ethereum | PoW | ✗ | $O(n)$ | I |
| BCoT Sentry, 2021 [15] | Ethereum | PoW | ✗ | $O(m*n)$ | I |
| BlockAuth, 2021 [16] | Hyperledger Fabric | PBFT | ✗ | $O(n^2)$ | I |
| SmartEdge- Ethereum, 2018 [17] | Ethereum | PoW | ✗ | $O(n)$ | A |
| DAMFA, 2020 [18] | Namecoin | PoW | ✗ | $O(n)$ | I |
| BCTrust, 2018 [19] | Ethereum | PoW | ✗ | $O(n)$ | CI |
| Blockchain-based User Authentication, 2018 [20] | Ethereum | PoW | ✗ | $O(n)$ | ✓ |
| WOT, 2017 [21] | Ethereum | PoW | ✗ | $O(n)$ | ✗ |
| Blockchain-Based IoT Authentication, 2021 [22] | Ethereum Hyperledger Fabric | PoW/PBFT | ✗ | $O(n)$ | I |
| Secure Combination of PUF-based NFT, 2021 [23] | Ethereum | PoW | ✓ | $O(n)$ | I |
| **DSCoT, 2022** | Hyperledger Besu | IBFT 2.0 | ✓ | $O(n)$ | ✓ |

### 4.4 Time Complexity for Latency

The proposed architecture does not modify data on the blockchain to verify the identity of all the functions and components. An innovative approach of *call* methods has been designed to query the smart contract for the status of assets in the NFT registry. It would not amend any data on the chain but will help save the transaction cost (Ether/Gewi) and be efficient in terms of time complexity.

- The "*adminAdd()*" call method has been designed to find the admins/Owner addresses, as shown in Figure 15 (A), which shows " 0x5B38Da6a701c568545dCfcB03FcB875f56beddC4" as an admin NFT-based EOA.
- The "*No_ofAdmins()*" call method has been designed to find the total number of admins/Owners as shown in Figure 15 (B), which shows "2" admin addresses exist.
- The "*user_Devices_Add()*" call method has been designed to find the total number of devices mapped to a specific user, as shown in Figure 15 (C), which shows that two NFT-based EOAs exist, i.e., fog: "0x78731D3Ca6b7E34aC0F824c42a7cC18A495cabaB" and IoT device:

"0x617F2E2fD72FD9D5503197092aC168c91465E7f2".

- The "*tokens_Issued()* call method has been designed to find a total number of NFTs, as shown in Figure 15 (D), which shows "0xf63fee14c773d0896382c7b8cd950adae380254bd7a346cb965818fab9143d82, 1657188740" generated NFT with a block timestamp.

Hyperledger Besu is an Ethereum-based private chain in which the time to generate new blocks depends on the block size. The transactions that cost transaction fees can have delays, but increasing transaction fees can solve the problem. We made more than 500 calls to the functions mentioned above, found it efficient, and did not find the transaction charging any gas fees. Assuming that there are 'n' IoT devices that require identity authentication, the proposed architecture presents $O(n)$ time complexity.

(A) adminAdd()

(B) No_ofAdmins()

(C) Users_devices()

(D) Tokens_Issued()

Figure 15. The time complexity of proposed DSCoT

## Conclusion

The cyber-physical systems in smart city architectures are prone to various adversaries. A distributed model based on blockchain tokenization has been proposed. A novel architecture of non-fungible tokens (NFTs) based on the ERC-721 standard for smart device representation and authentication mechanism of these smart devices together with the admin/owner and the user has been proposed to mitigate adversarial issues. An NFT-based smart contract has been developed using Remix IDE on a private blockchain utilizing a robust consensus mechanism, i.e., IBFT 2.0. The implementation was carried out for all the functions and procedures using NFT-based EOAs assigned to all the components in the smart contract. The security services, such as Confidentiality, Integrity, Availability (CIA), and authorization, were successfully deployed. Since NFTs are non-interchangeable and unique identifying codes representing each physical asset's ownership, a mechanism has been devised for user and device authentication. The evaluation of the proposed functions and components has been carried out in terms of Gas consumption, Efficiency, and Time complexity, showing promising results. Comparatively, the Gas consumption for minting DSCoT NFT showed approximately ≈ 27%, while a DSCoT *approve()* was approximately ≈ 11% more efficient, respectively. An innovative approach of *call()* methods has been designed to query the smart contract for the status of assets in the NFT registry. It does not amend any data on the chain, thus helping save the transaction cost (in Ether/Gewi) and making the proposed architecture efficient in terms of time complexity. In contrast, the smart devices' digital representation (IoT, fog) will be achieved using the proposed DSCoT NFTs. The architecture of this paper will deploy the use case scenarios for smart houses and smart hospitals in the future to further validate the security services and validation, and the obtained results will be reported accordingly.

## Acknowledgments

The author would like to express sincere appreciation for the resources provided by Universiti Brunei Darussalam (UBD), Brunei Darussalam.

CONFLICT OF INTEREST

The author(s) declared no potential conflicts of interest concerning this article's research, authorship, and/or publication.